\documentclass[graybox]{svmult}

\usepackage{type1cm}        
\usepackage{makeidx}         
\usepackage{graphicx}        
\usepackage{multicol}        
\usepackage[bottom]{footmisc}
\usepackage{amsmath, todonotes}
\usepackage[normalem]{ulem}
\usepackage{changes}

\makeindex             

\newcommand{\ie}{\textit{i.e.}}

\newcommand{\id}{\textrm{d}}

\def \dr{D_R}

\newcommand{\eref}[1]{Eq.~\eqref{#1}}

\def\bea{\begin{eqnarray}}
\def\eea{\end{eqnarray}}

\def\ba{\begin{array}}
\def\ea{\end{array}}
\def\n{\nonumber}

\def\la{\langle}
\def\ra{\rangle}

\begin{document}

\title*{Target search by active particles}
\author{Urna Basu, Sanjib Sabhapandit, Ion Santra}
\institute{Urna Basu
  \at S. N. Bose National Centre for Basic Sciences, Kolkata 700106, India \email{urna@bose.res.in}
\and Sanjib Sabhapandit \at  Raman Research Institute, Bengaluru 560080, India \email{sanjib@rri.res.in}
\and Ion Santra \at Institute for Theoretical Physics, Georg-August-Universit\"at  G\"ottingen, 37073 G\"ottingen, Germany, \email{ion.santra@theorie.physik.uni-goettingen.de}}
%
%
\maketitle

\abstract{Active particles, which are self-propelled nonequilibrium systems, are modelled by overdamped Langevin equations with colored noise, emulating the self-propulsion. In this chapter, we present a review of the theoretical results for the target search problem of these particles. We focus on three most well-known models, namely, run-and-tumble particles, active Brownian particles, and direction reversing active Brownian particles, which differ in their self-propulsion dynamics. For each of these models, we discuss the first-passage and survival probabilities in the presence of an absorbing target. We also discuss how resetting helps the active particles find targets in a finite time. }

\section{Introduction}
\label{sec:intro}

Survival of most species depends on successful search operations.
In the animal world, life begins with a successful search for a mate and thereupon sperm cells searching for an oocyte to fertilize~\cite{ikawa2010fertilization}. Similarly, in the plant kingdom, the propagation of species depends on pollination which is nothing but a random search process. 
Sustenance of animals requires foraging and finding suitable shelters~\cite{viswanathan2011physics}. Even within our body, the proteins use random search to locate their specific targets to bind on DNA~\cite{chen2014single}. Apart from these fundamental biological processes, search processes are also an indispensable part of human society---searching for information on the web, searching for lost items, trying to find the right word during a conversation are part of our daily lives. Random search optimization algorithms play a crucial role in computer science --- finding the ground state of a complex system, combinatorial optimization problems, cryptography, artificial intelligence applications such as machine learning, natural language processing all depend on randomized search algorithms.

Despite the diversity, a commonality unites most search problems; See the special issue Ref.~\cite{da2009random} for a comprehensive list of diverse search problems. Typically, the searcher `moves' stochastically, exploring the space and the process culminates upon finding the target. The underlying commonality allows us to use minimal stochastic models for such random search processes, the paradigmatic examples being the random walk and its continuous counterpart, the Brownian motion. Due to their inherent stochasticity, the methods of statistical physics are quite useful to study these random search models~\cite{Benichou2011,bray2013persistence}.

One of the most important observables, in the context of search processes, is the time required to find a target. Mathematically, the time at which a stochastic searcher hits a target for the first time is known as the first-passage time. This time $t$ is evidently a random quantity, whose distribution $F(a,t)$ also depends on the separation $a$ between the starting position and the location of the target. Another related quantity is the survival probability $Q(a,t)=\int_{t}^{\infty}\,F(a,t')\,dt'$, which denotes the probability that the searcher does not hit the target up to time $t$.
Therefore, the mean first-passage time, i.e., the average time to find the target, is given by,
\begin{equation}
    T(a)=\int_{0}^{\infty}t\, F(a,t)\,dt = \int_0^{\infty} Q(a,t)\,dt.
    \label{mfpt:gen}
\end{equation}

For a Brownian particle moving in one-dimension with diffusion constant $D$, the first-passage time distribution and the corresponding survival probability are given by,
\begin{equation}
F(a,t) = \frac{1}{\sqrt{4 \pi D}}\,\frac{a}{t^{3/2}}\, \exp\left(-\frac{a^2}{4 D t}\right)~~ \text{and} \quad Q(a,t) = \text{erf}\left(\frac{a}{ \sqrt{4Dt}}\right),
\label{FPP-Brownian}
\end{equation}
respectively~\cite{redner2001guide}. At large time, $Q(a,t)\sim t^{-\alpha}$ with $\alpha = 1/2$, where $\alpha$ is known as the persistence exponent.  For $N$ independent searchers $\alpha = N/2$, and consequently, $F_N(a,t)\sim t^{-(N/2+1)}$. Thus, the mean first-passage time is finite only if there are three or more searchers ($N\geq 3$).

It turns out that the searching process becomes drastically faster if the particle intermittently returns to the starting point and restarts the search~\cite{evans2011diffusion, evans2011diffusionprl}. Let $Q^{(r)}(a,t)$ denote the survival probability for a Brownian particle in the presence of such a stochastic resetting with a rate $r$. Its Laplace transform is related to the Laplace transform of $Q(a,t)$ by~\cite{evans2011diffusionprl}, 
\begin{align}
    \tilde Q^{(r)} (a,s)=\frac{ \tilde Q (a,r+s)}{1-r  \tilde Q (a,r+s)},
    \label{surv:diff-res}
\end{align}
where $\tilde Q^{*}(a,s) = \int_0^\infty Q^{*}(a,t)\, e^{-s t}\, dt$ denotes Laplace transform. The corresponding  mean first-passage time is nothing but, 
\begin{equation}
    T_r= \tilde Q^{(r)} (a,0)=\frac{\tilde Q (a,r)}{1-r  \tilde Q (a,r)}= -\frac{1}{r} + \frac{1}{r}\, \exp\left(a\,\sqrt{r/D}\right),
    \label{MFPT-diffusion-resetting}
\end{equation} 
which is finite for any non-zero resetting rate. It diverges for both small and large $r$: as a power-law $\sim r^{-1/2}$ as $r\to 0$, and stretched exponentially as $r\to\infty$, with a minimum at an optimal resetting rate satisfying  $r^*(a^2/D)=2.539 \cdots $.

The Brownian search is a Markov process. This is not true for most living organisms, which often have a finite memory, manifested in their persistent motion along their body axis. For example,  living organisms ranging from bacteria at the microscopic scale to an ant or the birds, at the macroscopic scale,  move persistently for a certain time before changing the direction substantially. These types of self-propelled motion are often referred to as \emph{active motion} and the agents displaying such motion are referred to as \emph{active particles}. Such motion has  also been realized in synthetic systems like Janus  particles~\cite{yang2012janus,RevModPhys.88.045006,walther2013janus}. For example, when a polystyrene spherical bead with one side coated with platinum is immersed in hydrogen peroxide solution, the platinum side  catalyzes the hydrogen peroxide to oxygen and water, causing a self-propulsion of the bead by diffusiophoresis~\cite{howse2007self}. These artificial active particles, including nano/microscopic machines, has a diverse range of potential applications in biomedical target search~\cite{nelson2010microrobots,wang2012nano}. 
Target search by such non-Markovian processes are expected to exhibit richer behaviour compared to the simple Brownian search, which is the subject of this chapter.

The self-propelled active motions, despite having complex and varied underlying mechanisms, can be broadly categorized into a few classes depending on their microscopic swimming strategies. In the minimal statistical models, the time-evolution of the position vector $\boldsymbol{r}(t)$ of a single active particle can be described by the overdamped Langevin equation, $\dot{ \boldsymbol{r}}(t)=v_0\,\hat{\boldsymbol n}(t)$, where the different swimming strategies are encoded by the different stochastic dynamics of the unit vector $\hat{\boldsymbol n}(t)$, indicating the orientation of the body axis. Perhaps the most well-known model is the so called run-and-tumble particle (RTP), where the active particle moves in a straight line before intermittently changing its direction $\hat{\boldsymbol{n}}$  in a discrete manner. The orientation $\hat{\boldsymbol{n}}$ undergoes a continuous change via a rotational diffusion for an active Brownian particle (ABP), whereas, a direction reversing active Brownian particle (DRABP) undergoes an intermittent complete reversal of direction in addition to the continuous ABP dynamics. The speed $v_0$ is constant in all these models.

In this Chapter, we discuss the survival probability and first-passage time distribution for RTP, ABP and DRABP in the presence of fixed targets. We further explore how stochastic resetting optimizes the search for these systems. We conclude with a summary and some open problems.

\section{Run-and-tumble particles}
\label{sec:rtp}

Run-and-tumble particle (RTP) is one of the first theoretical models of active particles, developed to mimic the overdamped motion of the {\it E. coli} bacteria~\cite{berg1972,lovely1975} that consists of alternating run and tumble phases. In the run phase, the bacterium moves almost in a straight line at a constant speed, whereas in the tumble phase, it reorients randomly, resulting in a change of the direction of motion. The typical duration of a tumbling phase is much shorter than that of a run phase. The RTP model assumes the tumbles to be instantaneous events occurring at a constant rate $\gamma$. Consequently, the position $\boldsymbol{r}$ and the unit vector $\hat{\boldsymbol n}$, indicating the direction of motion of an RTP evolve by,
\begin{align}
    \dot{ \boldsymbol{r}}(t)=v_0\,\hat{\boldsymbol n}(t),\quad\text{and}~~\hat{\boldsymbol n}\to\hat{\boldsymbol n}'~~\text{with rate}~\gamma,
    \label{eq:rtp:ddim}
\end{align}
where the speed $v_0$ remains constant during the run phases.
In other words, the duration $\tau$ of each run phase is drawn independently from the exponential distribution $\rho(\tau)=\gamma\,\exp(-\gamma \tau)$. Note that, while the time evolution of the position $ {\boldsymbol{r}}(t)$ is non-Markovian by itself, the combined $( {\boldsymbol{r}(t)},{\boldsymbol n}(t))$ process is Markov.

The first-passage time distribution has been computed exactly for a one-dimensional RTP in the presence of a fixed absorbing boundary~\cite{1drtp}. In one dimension, the unit vector $\hat{\boldsymbol n}$ reduces to a dichotomous random variable $\sigma=\pm 1$, and \eref{eq:rtp:ddim} simplifies to,
\begin{align}
\dot{x}(t)=v_0\sigma(t),
\label{eq:rtp:eom}
\end{align}
where $\sigma\to-\sigma$ at a rate $\gamma$. In the following, we present the main steps of the derivation of the survival probability, which was obtained in~\cite{1drtp}.

\subsection{Survival Probability}
 The survival probability $Q_{\sigma}(x,t)$, which denotes the probability that the RTP has not crossed the absorbing boundary at the origin up to time $t$, starting from an initial position $x>0$ and orientation $\sigma$, follows a backward Fokker-Planck equation~\cite{risken2012fokker}. To obtain the BFPE it is convenient to consider a time-interval $[0,t+\Delta t]$, separated into two parts $[0,\Delta t]$ and $[\Delta t,t+\Delta t]$. Using the temporal homogeneity of the $(x,\sigma)$ process and the fact that in the first interval, either $\sigma \to -\sigma$ with probability $\gamma\Delta t$, or the particle moves a distance $\sigma v_0 \Delta t$ ballistically with probability $(1-\gamma\Delta t)$, we have, 
\begin{math}
Q_{\sigma}(x,t+\Delta t)=(1-\gamma\Delta t) Q_{\sigma}(x+v_0\sigma\Delta t,t)+\gamma\Delta t Q_{-\sigma}(x,t).
\end{math}

Taking a Taylor series expansion and taking the $\Delta t\to 0 $ limit, we get the backward Fokker-Planck equation~\cite{risken2012fokker},
\begin{align}
\frac{\partial Q_{\sigma}(x,t)}{\partial t}=v_0\sigma\frac{\partial Q_{\sigma}(x,t)}{\partial x}-\gamma\bigg[Q_{\sigma}(x,t)-Q_{-\sigma}(x,t)\bigg].
\label{bfp:rtp1d}
\end{align}
which have the initial conditions $Q_{\sigma}(x,0)=1$. Since a particle starting at infinity cannot cross the origin at any finite time, we have the boundary condition $Q_{\sigma}(x \to \infty,t)=1$ for all times. An RTP starting with a negative velocity at the origin gets immediately absorbed, leading to the boundary condition $Q_{-}(0,t)=0$.  On the other hand, an RTP starting with a positive velocity does not get absorbed immediately and therefore the boundary condition on $Q_{+}(0,t)$ can not be specified and has to be obtained from the above equation at $x=0$, using $Q_{-}(0,t)=0$.

Taking a Laplace transform, defined by $\tilde{Q}_{\sigma}(s)=\int_0^{\infty}dt~e^{-st}Q_{\sigma}(t)$,
\begin{align}
\mathcal{L}_{\sigma}\tilde Q_{\sigma}(x,s) = 1+\gamma \tilde Q_{-\sigma}(x,s)\quad\text{with}~~\mathcal{L}_{\sigma}\equiv s+\gamma-v_0\sigma\frac{\partial}{\partial x},
\end{align}
where the boundary conditions are given by $\tilde Q_{\sigma}(x\to\infty,s)=1/s$, $\tilde Q_{-}(0,s)=0$ and $\mathcal{L}_{+}\tilde Q_{+}(0,s)=1$. 
Operating $\mathcal{L}_{\sigma}$ on both sides of the above equation, we get,
\begin{align}
    \Big[v_0^2\frac{\partial^2}{\partial x^2}-s(s+2\gamma )\Big]Q_{\sigma}(x,s)=-(s+2\gamma).
\end{align}
Solving the above equation and using the boundary conditions, we get, 
\begin{align}
\tilde Q_{+}(x,s)&=\frac{1}{s}\left(1-\frac{(s+\gamma)-\lambda}{\gamma}e^{-\lambda x/v_0}\right),~~ 
\tilde Q_{-}(x,s)=\frac{1}{s}(1-e^{-\lambda x/v_0})
\label{eq:1drtpQ-},
\end{align}
where $\lambda=\sqrt{s(s+2\gamma)}$. It is immediately evident from the above equations that the mean first-passage times, $T_{\pm}(x)=\tilde Q_{\pm}(x,0)$ is infinite. 

Since the Laplace transforms of the first-passage time distribution $\tilde F_{\pm}(x,s)=\int_0^\infty F_\pm(x,t)\, e^{-st}\, dt$ are related to the  survival probability  by $\tilde Q_{\pm}(x,s)=[1-\tilde F_{\pm}(x,s)]/s$, the Laplace transforms $\tilde F_{\pm}(x,s)$ are readily available from the above equations. Inverting the Laplace transform yields,
  \begin{align}
   F_+(x,t)&=\frac{\gamma e^{-\gamma t}}{t+x/v_0}\Bigg[\frac{x}{v_0} I_0\left(\gamma \sqrt{t^2-(x/v_0)^2}\right)\cr  &+\frac{1}{\gamma}\frac{\sqrt{t-x/v_0}}{\sqrt{t+x/v_0}}I_1\left(\gamma \sqrt{t^2-(x/v_0)^2}\right)\Bigg]\Theta(t-x/v_0), \label{eq:Fplus}\\
    F_-(x,t)&=\frac{\gamma e^{-\gamma t}\,x/v_0}{\sqrt{t^2-\left(x/v_0\right)^2}}I_1\left(\gamma \sqrt{t^2-\left(x/v_0\right)^2}\right)\Theta(t-x/v_0)\cr 
    & + e^{-\gamma t}\delta\left(t-x/v_0\right). \label{eq:Fminus}
 \end{align}
Note that, the minimum possible value of the first passage time is  $t=x/v_0$ which is achieved  when the particle starts with a negative velocity, i.e., $\sigma=-1$ does not tumble before reaching the target. The Dirac-delta function appearing in $F_-(x,t)$ corresponds to these deterministic trajectories which occur with a probability $e^{-\gamma t}$. There is no analogous  term for $F_+(x,t)$ as a particle starting with a positive velocity, i.e., $\sigma=1$, must undergo at least one tumbling in order to reach the target at the origin. For all the trajectories undergoing at least one tumbling event, the first passage time $t > x/v_0$. The Heaviside Theta function $\theta(t-x/v_0)$ appearing 
in Eqs.~\eqref{eq:Fplus} and \eqref{eq:Fminus} indicate the contributions from these trajectories.

Evidently, the above expressions for the first-passage time distribution for an RTP is very different from that of a Brownian particle. However, in the limit $\gamma\to\infty$ and $v_0\to\infty$, while keeping $v_0^2/\gamma= 2 D_{\text{RT}}$ finite, the above equations reduce to the first-passage time distribution for the Brownian motion given by \eqref{FPP-Brownian} with $D_{\text{RT}}$ being the diffusion coefficient.

For any finite $\gamma$ and $v_0$, the large-time $t\gg x/v_0,\,\gamma^{-1}$ asymptotic behavior comes out to be,
 \begin{align}
 F_{-}(x,t)\simeq\frac{\sqrt{\gamma}/v_0}{\sqrt{2\pi }}\frac{x}{t^{3/2}},~~F_{+}(x,t)\simeq\frac{\sqrt{\gamma}/v_0}{\sqrt{2\pi }}\,\frac{\bar x}{ t^{3/2}}
 \label{1drtp-larget}
\end{align}  
where $\bar x=x+v_0/\gamma$. A particle starting with a positive velocity i.e., $\sigma=1$ typically moves a distance $v_0/\gamma$ before the first tumbling event. Since the particle moves with a negative velocity immediately after the first tumbling event, the first-passage time distribution $F_{+}(x,t)$ is equivalent to $F_{-}(\bar x,t)$ for large $t$.

Of particular interest is the special case $x=0^+$, i.e.,  when the initial position of the particle is infinitesimally close to the target. Obviously, if the initial velocity is negative i.e., $\sigma=-1$, then the particle gets absorbed immediately. Consequently,  $F_-(0,t)=\delta(t)$, which is also obtained from \eref{eq:Fminus} in the $x=0$ limit. The corresponding survival probability $Q_-(0,t)=0$, which is, in fact, one of the boundary conditions on \eref{bfp:rtp1d}. 
On the other hand, a particle starting from $x=0^+$ with a positive velocity still survives for a finite time, characterized by the distribution $ F_+(0,t)=t^{-1}e^{-\gamma t}I_1(\gamma t)$ obtained from \eref{eq:Fplus} by setting $x=0$.
The corresponding survival probability in this case, is given by, $Q_+(0,t)=e^{-\gamma t}\bigl[I_0(\gamma t)+I_1(\gamma t)\bigr]$. Therefore choosing the initial velocity direction $\sigma=\pm 1$ with equal probability $1/2$, we get the total survival probability $Q(0,t) = [Q_+(0,t)+ Q_-(0,t)]/2$ as, 
\begin{align}
    Q(0,t)=\frac{1}{2}e^{-\gamma t}\bigl[I_0(\gamma t)+I_1(\gamma t)\bigr].
    \label{eq:sp-1drtp}
\end{align}

In the one-dimensional model of RTP discussed so far, the velocity direction $\sigma=\pm 1$ switches at each tumbling event, which occurs with a rate $\gamma$. One can also think of a different version of an RTP, where at each tumbling event occurring with rate $\gamma$, the new velocity direction is chosen independently from  $\sigma=\pm 1$ with equal probability $1/2$. Therefore, in a small time interval $dt$, the probability that the RTP reverses its velocity direction is $\gamma dt/2$, resulting in a waiting time distribution $\rho(\tau)=(\gamma/2)\,e^{-\gamma \tau/2}$. For such an RTP, the survival probability \eref{eq:sp-1drtp} becomes,
\begin{align}
    Q(0,t)=\frac{1}{2}e^{-\gamma t/2}\bigl[I_0(\gamma t/2)+I_1(\gamma t/2)\bigr].
    \label{eq:sft-rtp-universal}
\end{align}
Note that the constant speed $v_0$ can be thought of as drawing the speed from a distribution $W(v)= \delta(v-v_0)$ after each tumbling event. 

It turns out that \eref{eq:sft-rtp-universal} is universal~\cite{Mori2020} in the sense that it remains valid for a much wider class of RTP with any arbitrary speed distribution $W(v)$, as a consequence of the  Sparre Andersen theorem for one-dimensional  discrete-time random walks with  symmetric and continuous jump distributions~\cite{andersen1954fluctuations}. Naturally, this generalized RTP also describes the dynamics of the $x$ component of a $d$-dimensional RTP, and \eref{eq:sft-rtp-universal} remains valid for the corresponding survival probability~\cite{Mori2020}.

\subsection{RTP with Resetting}
The presence of stochastic resetting expedites the target search process for an RTP, similar to a Brownian particle. In this section, we discuss the behavior of the first-passage properties of an RTP starting from an initial position $x_0>0$, in the presence of an absorbing boundary at $x=0,$ undergoing stochastic resetting with rate $r$. Since the state space of RTP consists of $(x,\sigma)$, we need to specify the resetting protocol for both $x$ and $\sigma$. Here we consider the resetting protocol, where at each resetting event, the position $x(t)\to x_0$ and the velocity direction $\sigma(t)\equiv\sigma'\to\sigma''$ with probability $p_{\sigma',\sigma''}$.


We assume that the RTP starts with the equilibrium state of the velocity orientation, where $\sigma=\pm1$ with equal probability $1/2$. Our goal is to find the corresponding survival probability $Q^{(r)}(x_0,t)$, ---i.e., the probability that starting from the position $x_0$, the RTP has not crossed the target up to time $t$. To this end, it is useful to start with a last renewal equation for $Q^{(r)}_{\sigma',\sigma}(x_0,t)$, which denotes the probability that starting from  the state $\{x(0)=x_0, \sigma(0)=\sigma\}$, the RTP has not crossed the target (absorbing boundary) up to time $t$ and $\sigma(t)=\sigma'$. Evidently, 
\begin{align}
    Q^{(r)}(x_0,t)=\frac{1}{2}\sum_{\sigma,\sigma'}Q^{(r)}_{\sigma',\sigma}(x_0,t)
    \label{rtp:survsum}
\end{align}

To construct the renewal equation for $Q^{(r)}_{\sigma',\sigma}(x_0,t)$, we consider two scenarios: (i) there is no resetting during interval $(0,t)$, which has a probability $e^{-rt}$, and (ii) there is at least one resetting and the last reset occurs within the time interval $(t-\tau,t-\tau+d\tau)$ with a probability $rd\tau$. The probability that there is no reset during the remaining time is $e^{-r\tau}$.
Suppose $\sigma(t-\tau)=\sigma''$ and during the reset $\sigma'' \to \sigma'''$ with probability $p_{\sigma'',\sigma'''}$. Summing over all possible $\sigma''$, we get the renewal equation as,
\begin{align}
Q^{(r)}_{\sigma',\sigma}&(x_0,t)=e^{-r t}\,Q_{\sigma',\sigma}(x_0,t)\cr
&+r\int_0^t d\tau \, e^{-r\tau} \sum_{\sigma'',\sigma'''}Q_{\sigma',\sigma'''}(x_0,\tau)\,    Q^{(r)}_{\sigma'',\sigma}(x_0,t-\tau)\, p_{\sigma'',\sigma'''},
\label{RTP:renewal}
\end{align}
where $Q_{\sigma',\sigma}(x_0,t)\equiv Q^{(0)}_{\sigma',\sigma}(x_0,t)$ denotes the survival probability in the absence of resetting. Taking a Laplace transform in the above equation,  $\tilde Q^{*}_{\sigma',\sigma}(x_0,s)= \int_0^\infty Q^{*}_{\sigma',\sigma}(x_0,t)\, e^{-st}\, dt$, we get, 
\begin{equation}
    \tilde Q^{(r)}_{\sigma',\sigma}(x_0,s)=\tilde Q_{\sigma',\sigma}(x_0,r+s)+r \sum_{\sigma'',\sigma'''}\tilde Q_{\sigma',\sigma'''}(x_0,r+s)\,  \tilde Q^{(r)}_{\sigma'',\sigma}(x_0,s)\, p_{\sigma'',\sigma'''}.
\end{equation}
This is a system of four linear equations and can be solved easily to obtain $\tilde Q^{(r)}_{\sigma',\sigma}(x_0,s)$ in terms of $\tilde Q_{\sigma',\sigma}(x_0,s)$. Summing over $\{\sigma,\sigma'\}$, {\it \`a la} \eref{rtp:survsum}, we get, 
\begin{align}
   \tilde{Q}^{(r)}(x_0,s)=\frac{\displaystyle \tilde Q(x_0,r+s)+\frac{r}{2}\left(\sum_{\sigma,\sigma'}\sigma\sigma'\, p_{\sigma,\sigma'}\right)\mathbf{det}[\tilde {Q}(x_0,r+s)]}{\displaystyle 1-r\sum_{\sigma,\sigma'}\, p_{\sigma,\sigma'}\,\tilde Q_{\sigma,\sigma'}(x_0,r+s)+r^2\mathbf{det}[p]\times \mathbf{det}[\tilde {Q}(x_0,r+s)]},
   \label{rt:reset:gen}
\end{align}
where, $[p]$ and $[\tilde Q]$ denote $2\times 2$ matrices, given by,
\begin{equation}[p]=\begin{bmatrix}
    p_{+,+}&p_{+,-}\\
    p_{-,+}&p_{-,-}
\end{bmatrix}\quad\text{and}\quad
[\tilde Q]=\begin{bmatrix}
   \tilde Q_{+,+} & \tilde Q_{+,-}\\
   \tilde Q_{-,+} & \tilde Q_{-,-}
\end{bmatrix}.
\end{equation} 
 It is noteworthy that \eref{rt:reset:gen} is very different from its passive counterpart given by \eref{surv:diff-res} and depends explicitly on how the active degree of freedom is reset. Evidently, the mean first-passage time can be obtained by setting $s=0$ in \eref{rt:reset:gen},
\begin{equation}
    T^{(r)}(x_0)=\frac{\displaystyle \tilde Q(x_0,r)+\frac{r}{2}\Big(\sum_{\sigma,\sigma'}\sigma\sigma'\, p_{\sigma,\sigma'}\Big)\mathbf{det}[\tilde {Q}(x_0,r)]}{\displaystyle 1-r\sum_{\sigma,\sigma'}\, p_{\sigma,\sigma'}\,\tilde Q_{\sigma,\sigma'}(x_0,r)+r^2\mathbf{det}[p]\times \mathbf{det}[\tilde {Q}(x_0,r)]}.
    \label{rtp:reset:mfpt:gen}
\end{equation}

Note that, the presence of multiple resetting options for the orientation of an RTP makes the above expression of the MFPT quite different from its Brownian counterpart, given by a much simpler form in \eref{MFPT-diffusion-resetting}.

The Laplace transforms $\tilde Q_{\sigma,\sigma'}(x_0,s)$ can be obtained using a backward Fokker-Planck equation approach~\cite{evans2018run}, similar to the one discussed in the previous section,
\begin{equation}
\begin{split}
    \tilde Q_{+,+}(x_0,s)&=\frac{1}{s(s+2\gamma)}\left[s+\gamma-(s+\gamma-\lambda)e^{-\lambda x_0/v_0}\right],\cr
    \tilde Q_{+,-}(x_0,s)&=\frac{\gamma}{s(s+2\gamma)}\left[ 1-e^{-\lambda x_0/v_0}\right],\cr
    \tilde Q_{-,+}(x_0,s)&=\frac{1}{s(s+2\gamma)}\left[\gamma-\frac{(\gamma+s)(\gamma+s+\lambda)}{\gamma}e^{-\lambda x_0/v_0}\right],\cr
    \tilde Q_{-,-}(x_0,s)&=\frac{s+\gamma}{s(s+2\gamma)}\left[1-e^{-\lambda x_0/v_0} \right].
    \end{split}
    \label{rtp:surv-noreset}
\end{equation}

\begin{figure}[h]
    \centering
    \includegraphics[width=\hsize]{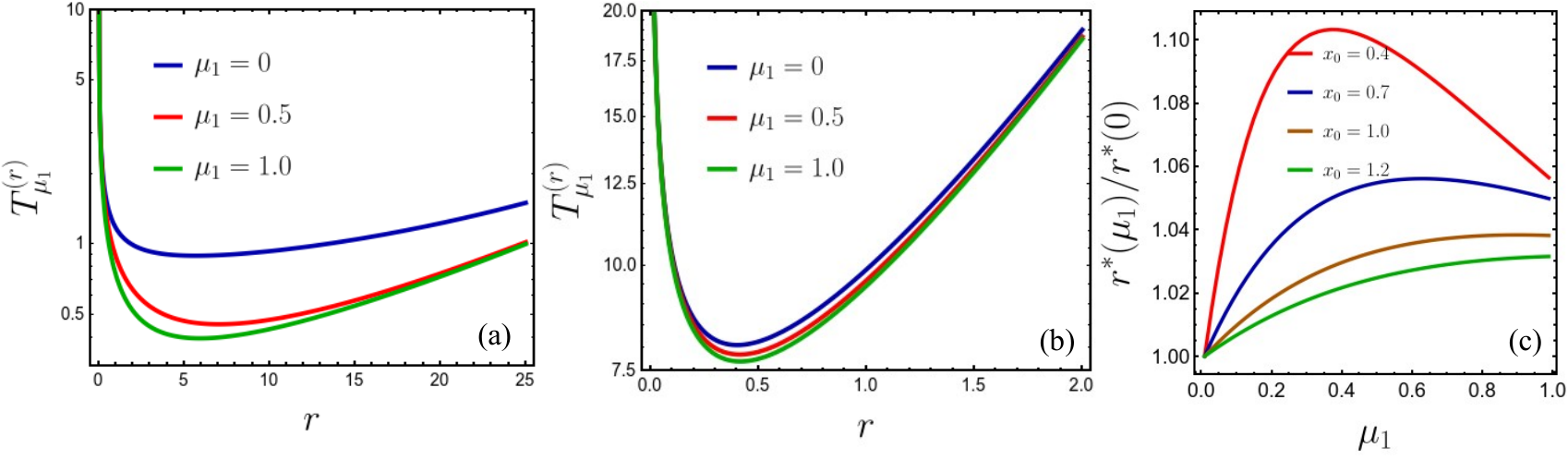}
    \caption{Plots for protocol I with $\gamma=1$ and $v_0=1$. (a) shows MFPT vs $r$ for $x_0=0.1<\ell_a=1$; (b) shows MFPT vs $r$ for $x_0=1.1>\ell_a$; (c) shows the normalized optimal resetting rate as a function of $\mu_1$, which shows a non-monotonic behavior for $x_0<\ell_a$ and a monotonically decreasing behavior for $x_0>\ell_a$.}
    \label{fig:mfpt1}
\end{figure}

In the following, we consider two distinct resetting protocols for $\sigma$.  In the first case~\cite{evans2018run}, $\sigma$ either reverses sign with probability $\mu_1$ or remains unchanged  with probability $1-\mu_1$ at reset, where $0\leq\mu_1\leq 1$. This corresponds to $p_{\sigma,\sigma'}=\mu_1 \delta_{\sigma',-\sigma}+(1-\mu_1)\delta_{\sigma',\sigma}$, and the mean first-passage time, given by \eref{rtp:reset:mfpt:gen}, simplifies to,
\begin{equation}
    T_{\mu_1}^{(r)}=-\frac{1}{r}+\frac{1}{r+2\gamma-\kappa}\left[ \frac{2\gamma }{r}e^{\kappa x_0/v_0}-\frac{(2\mu_1-1)(\kappa-r)}{2(\mu_1 r+\gamma)}\right], \label{eq:RTP_T_P1}
\end{equation}
with $\kappa=\sqrt{r(r+2\gamma)}$. The mean first passage time shows a qualitatively different behavior depending on whether $x_0$ is smaller or larger than the persistence length scale, $\ell=v_0\gamma^{-1}$, of the RTP, as shown in Figs.~\ref{fig:mfpt1}(a) and (b), respectively. Though in both the cases, $T^{(r)}_{\mu_1}$ is finite for all $r>0$ and has a minimum at an optimum value $r^*$, the optimal resetting rate $r^*$ is a non-monotonic function of $\mu_1$ for $x_0<\ell_a$, while it decreases  monotonically with $\mu_1$ for $x_0>\ell_a$ [see Fig.~\ref{fig:mfpt1}(c)].

\begin{figure}[h]
    \centering
    \includegraphics[width=\hsize]{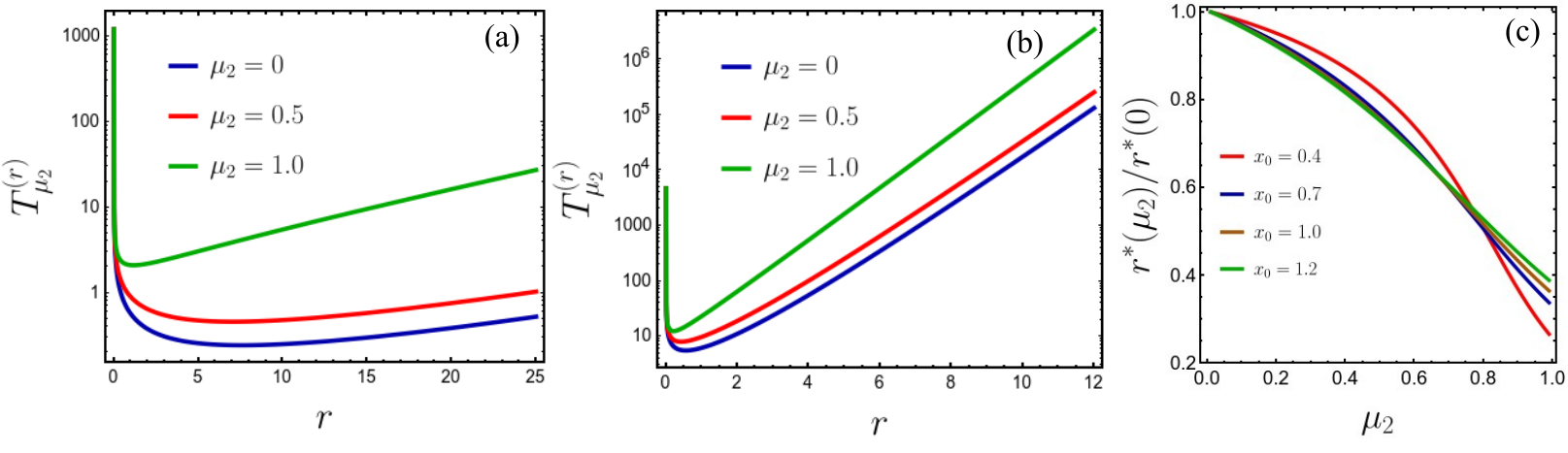}
    \caption{Plots for protocol II with $\gamma=1$ and $v_0=1$. (a) shows MFPT vs $r$ for $x_0=0.1<\ell_a=1$; (b) shows MFPT vs $r$ for $x_0=1.1>\ell_a$; (c) shows the normalized optimal resetting rate as a function of $\mu_2$, which always shows a monotonically decreasing behavior.}
    \label{fig:mfpt2}
\end{figure}

On the other hand, for the second case, a new velocity direction is chosen from $[1,-1]$ with a certain probability independent of the previous orientation. This corresponds to $p_{\sigma,\sigma'}=\mu_2 \delta_{\sigma,1}+(1-\mu_2)\delta_{\sigma',-1}$, where the mean first-passage time is given by,
\begin{align}
    T_{\mu_2}^{(r)}=-\frac{1}{r}+\frac{2\gamma\, e^{\kappa x_0/v_0}- (2\mu_2 -1)(\kappa -r)}{2 r[\gamma - \mu_2 (\kappa-r)]}. \label{eq:RTP_T_P2}
\end{align}
 The mean first-passage time $T^{(r)}_{\mu_2}$ is finite for all $r>0$ and has a minimum at an optimum value $r^*(\mu_2)$, which is shown in Fig.~\ref{fig:mfpt2}(a) and (b) for $x_0<\ell_a$ and $x_0>\ell_a$, respectively. The optimal resetting rate $r^*(\mu_2)$ decreases monotonically with $\mu_2$ [see Fig.~\ref{fig:mfpt2}(c)]. 
  In the Brownian limit $\gamma \to \infty, v_0 \to \infty$ with $v_0^2/\gamma=2D_{RT}$, both \eref{eq:RTP_T_P1} and \eref{eq:RTP_T_P2} converge to \eref{MFPT-diffusion-resetting}.

Note that, the two resetting protocols become equivalent for $\mu_1=\mu_2=1/2$ for our chosen initial condition. This special case corresponds to a complete resetting protocol, with both the position and the orientation being reset to their initial conditions. In fact, in this case, $p_{\sigma,\sigma'}=1/2$, for all $(\sigma,\sigma')$, and \eref{rtp:reset:mfpt:gen} becomes identical to the first equality of \eref{surv:diff-res}.

Several variants of this one dimensional run-and-tumble model have also been studied which show interesting first-passage properties. For example, exact results for the mean first-passage time for an RTP with space dependent velocity and tumbling rate, in the presence of external potential on an interval was obtained in~\cite{angelani2014first}. 
Survival probability has also been obtained for run-and-tumble motion  with position and orientation dependent tumbling rates~\cite{Singh_2020} and with external confining potential~\cite{dhar2019run}. First-passage behavior of RTPs in the presence of sticky boundaries have been studied using encounter based models~\cite{bressloff1,bressloff2}. Recently, it was shown that the mean first-passage time of a chiral RTP, with discrete orientation space, moving on a finite two-dimensional region, exhibits a minimum at an optimum value of chirality~\cite{mallikarjun2023chiral}. 
Effect of initial condition on the mean first-passage time of an RTP in the presence of two absorbing boundaries has been explored recently, using a perturbative analysis, in the small P\'eclet limit~\cite{iyaniwura2023asymptotic}. The first-passage properties of an RTP, with discrete internal states and moving on a lattice, was studied in~\cite{jose2022first} and it was shown that activity facilitates the return to the origin.
The effect of resetting on the mean first-passage time of an RTP moving diffusively in an anharmonic external potential has been studied in~\cite{scacchi2018mean}.

\section{Active Brownian particles}
\label{sec:abp}

Not all microorganisms exhibit a run-and-tumble motion. For example, some mutants of {\it E. coli} (CheC497), unlike a wild type, changes its direction of motion continuously~\cite{berg1972}. A similar motion is also observed for typical Janus particles~\cite{howse2007self}. An `active Brownian particle' models such motion, where the orientation vector of the particle undergoes a rotational Brownian motion on a unit sphere. The two-dimensional model is the simplest case, where the orientation vector $\hat{\mathbf{n}}=(\cos\theta,\sin\theta)$ and the position $\mathbf{r}(t)=(x(t),y(t))$ evolve according to,
\begin{align}
    \dot{x}(t) &=v_0 \cos \theta(t), ~~ \dot{y}(t)= v_0 \sin \theta(t), \quad\text{and}~~ \dot\theta(t) = \sqrt{2D_R}\, \zeta(t),
    \label{abp:langevin} 
   \end{align}
    where $\zeta(t)$ is a Gaussian white noise with zero mean and correlations $\la \eta(t)\eta(t')\ra=\delta(t-t')$.
    The rotational diffusion constant $D_R$ sets the active time scale $D_R^{-1}$, separating the short and long time dynamical regimes. The presence of activity leads to the emergence of a strongly anisotropic and nondiffusive behavior at the short-time regime $t\ll D_R^{-1}$.
 On the other hand, in the long-time regime  $t\gg D_R^{-1}$, the dynamics \eref{abp:langevin}  converges to a Brownian motion with an effective diffusion coefficient $D_{\textbf{{eff}}}=D_{\textbf{{AB}}}=v_0^2/(2D_R)$~\cite{Santra2022universal}. 
 Naturally, the first-passage properties also show different behaviors in the short-time and long-time regimes, which we discuss below.

\subsection{Survival probability}

The active nature of the short-time dynamics of ABP leads to strongly anisotropic first-passage properties. These effects are best understood if one looks at the survival probabilities along and orthogonal to the initial orientation of the ABP~\cite{Basu2018}. In the following we choose the initial orientation $\theta(0)=0$, and look at the survival probabilities along $x$ and $y$ directions (parallel and orthogonal to the initial orientation, respectively) separately. 

Let $Q_x(x_0,t)$ denote the probability that the ABP, starting from the $x=x_0$ line, has not crossed the absorbing boundary at $x=0$. Similarly, $Q_y(y_0,t)$ denotes the survival probability for the $y$-component, starting from the $y=y_0$ line with an absorbing boundary at $y=0$. Although it is straightforward to write the corresponding backward Fokker-Planck equations, it is hard to solve them analytically with appropriate  boundary conditions.
However, one can predict the leading order behaviors of $Q_x(x_0,t)$ and $Q_y(y_0,t)$  by analyzing the \eref{abp:langevin} in the short-time and long-time limits separately.

Since at late times $t \gg D_R^{-1}$, the active Brownian motion converges to a usual Brownian motion, one expects,
\begin{align}
    Q_x(x_0,t) \sim \frac 1{\sqrt{4 D_\text{eff}t}}, \quad Q_y(y_0,t) \sim \frac 1{\sqrt{4 D_\text{eff}t}}. \label{eq:Q_diff}
\end{align}
On the other hand, starting from $\theta(0)=0$, at short times $t\ll D_R^{-1}$,  the orientation angle $\theta(t) \sim O(\sqrt{D_R t}) \ll 1$. Consequently,  $\cos \theta(t) \simeq 1$ and $\sin \theta(t) \simeq \theta(t)$. The motion of the ABP in the short-time regime can then be described by the effective Langevin equations,
\begin{align}
    \dot x \simeq v_0, ~~ \dot y \simeq v_0 \, \theta(t), \label{eq:ABP_shortt}
\end{align}
where $\theta(t)$ undergoes a Brownian motion with diffusion constant $D_R$. 

The above effective dynamics implies that, in this short-time regime, $x(t)$ stays positive with probability close to unity, implying $Q_x(x_0,t) \simeq 1$. In the $x_0 \to 0$ limit, the crossover from this ballistic behaviour to the diffusive behaviour \eref{eq:Q_diff}, can be captured by the scaling form, 
\bea 
Q_x(x_0 \to 0,t) = {\cal F}_{x}\left( t \, D_R \right), \label{eq:Sx_scaling}
\eea
where the scaling function,
\begin{align}
{\cal F}_{x}(u) \sim 
\begin{cases}
1 &\text{for} \quad  u \ll 1, \cr 
u^{-1/2} &  \text{for} \quad u \gg 1. 
\end{cases}
\end{align}

\begin{figure}[t]
    \centering
    \includegraphics[width=10 cm]{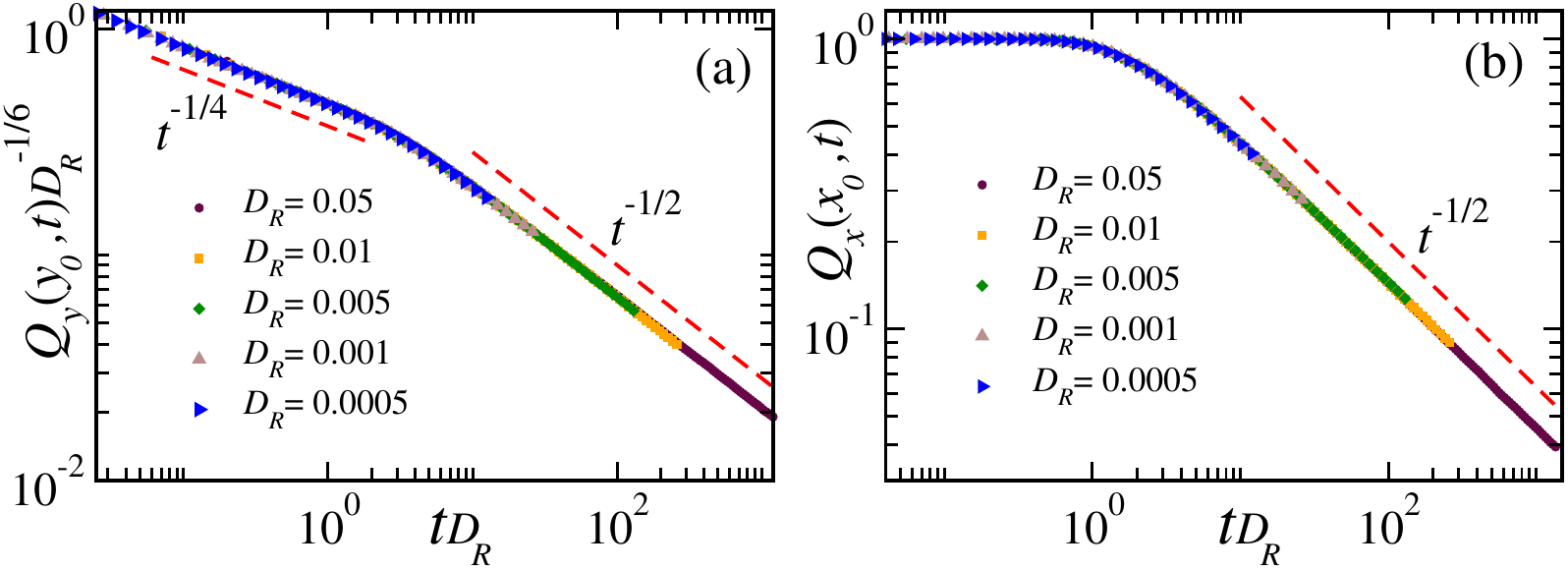}
    \caption{Survival probability of ABP: (a) Scaling collapse of $Q_y(y_0,t)$ according to \eref{eq:Sy_scaling} for different values of $D_R$, using data obtained from numerical simulations. The dashed lines indicate the analytically predicted behaviours for the short-time and long-time regimes. Here we have used $v_0=1$ and $y_0=0.1$. (b) Scaling collapse of $Q_x(x_0,t)$ according to \eref{eq:Sx_scaling} for different values of $D_R$, using data obtained from numerical simulations. The dashed line indicates the diffusive $t^{-1/2}$ decay. Here  $v_0=1$ and $x_0=0.1$. }
    \label{fig:abp_Qxy}
\end{figure}

To extract the short-time behaviour of the survival probability $Q_y(y_0,t)$ it is convenient to take another time derivative of $\dot y(t)$ in \eref{eq:ABP_shortt}, which yields,
\bea
\ddot y(t) \approx \sqrt{2\, v_0^2 D_R} \,\, \zeta(t),  \label{eq:RAM}
\eea
and $\zeta(t)$ is the white noise defined in \eref{abp:langevin}. 
The above equation is noting but the Langevin equation for the well-known `Random acceleration process' (RAP), which is one of the simplest non-Markovian processes [refs]. The first-passage properties of the RAP has been studied extensively in the literature, and the survival probability has been computed in the $y_0 \to 0$ limit [refs]. This result, translated to the case of short-time regime of ABP, gives,
\bea
Q_y(y_0,t) \simeq \frac {2^{ 5/6}\Gamma(-4/3)}{3^{2/3}\pi \Gamma(3/4)} \left(\frac{y_0\, D_R }{v_0}\right)^{1/6} \left(t\,D_R \right)^{-1/4}  \;.\label{eq:Sy_RAP}
\eea
Clearly, the above result implies that the $y$ component, i.e., the component orthogonal to the initial orientation, of the ABP shows a persistence exponent $\alpha_y = 1/4$ in the short-time regime. 
The crossover from this anomalous persistence behaviour to the diffusive behaviour \eref{eq:Q_diff} at late times suggests the following scaling form,
\bea
Q_{y}(y_0,t) = \left(\frac{y_0 \, D_R}{v_0}\right)^{1/6} {\cal F}_{y}\left(t\, D_R \right). \label{eq:Sy_scaling}
\eea
The scaling function ${\cal F}_{y}(u)$ has the limiting behaviors,
\begin{align}
{\cal F}_{y}(u) \sim 
\begin{cases}
u^{-1/4} \quad \text{for} \quad  u \ll 1 \cr 
u^{-1/2} \quad \text{for} \quad u \gg 1. 
\end{cases}
\end{align}
 Note that, due to the $t^{-1/2}$ tail of the survival probability, the mean first-passage time for both the $x$ and $y$ components of ABP diverge, similar to the Brownian motion. Figure \ref{fig:abp_Qxy} illustrates the scaling behaviors for $Q_x(x_0,t)$ and $Q_y(y_0,t)$ predicted in  \eref{eq:Sx_scaling} and  \eref{eq:Sy_scaling} which show the expected crossovers near $D_R t\sim 1$.

\subsection{ABP with Resetting}

Similar to the RTP, the presence of resetting also helps the ABP to find the target in a finite time. Since, for an ABP, the motion orthogonal to the initial orientation shows a much richer behaviour compared to the motion along the initial orientation, we concentrate on the first-passage properties in the presence of resetting for the orthogonal component only. The simplest resetting protocol is where the position and orientation are both reset to their initial values. As before, we set the initial orientation $\theta(0)=0$, so that the $y$-component represents the orthogonal motion. For this resetting protocol with a constant resetting rate $r$, and resetting position $y=y_0$, the last renewal equation takes the form, 
\begin{align}
    Q^{(r)}(y_0,t)=e^{-rt}Q(y_0,t)+r\int_0^{t}d\tau e^{-r\tau}Q^{(r)}(y_0,t-\tau)Q(y_0,\tau).
    \label{abp:renewal}
\end{align}
where  $Q^{(r)}(y_0,t)$ denotes the survival probability of the $y$ component in the presence of an absorbing wall at $y=0$.  Taking a Laplace transform of the above equation, we get the mean-first passage time, 
\bea 
T_y = \frac{\tilde Q(y_0,r)}{1- r \tilde Q(y_0,r)}.
\label{abp:mfpt}
\eea 
where $\tilde Q(y_0,r)$ denotes the Laplace transform of $Q(y_0,t).$  In general it is hard to calculate $T_y$ explicitly. However, for $D_R \ll r$ we can use the effective RAP picture to compute the MFPT. For an RAP starting at resy from a position $y_0$, the Laplace transform of the survival probability $\tilde Q(y_0,s)$, for an absorbing boundary at the origin is known exactly~\cite{burkhardt1993semiflexible},
\bea 
 \tilde Q(y_0,s) =\frac{1}{s}\left(1- p\left(\frac{y_0 r^{3/2}}{v_0 \sqrt{D_R}}\right)\right),
\eea
where,
\bea
p(w) = \int_0^\infty \frac{dz}{z^{5/3}} e^{-w z}\text{Ai}(z^{-2/3})\left[1+ \frac 1{4 \sqrt{\pi}} \Gamma\left(-\frac 12, \frac 2{3z} \right)\right].
\label{p(w)-ABP-reset}
\eea 
Thus for an ABP, being
reset to initial position $y_0$, with velocity $\theta=0$ at rate $r$, the mean first passage, using \eref{abp:mfpt} is given by,
\begin{align}
    T_y = \left(\frac{y_0}{v_0\sqrt{D_R}}\right)^{2/3}\mathcal{H}\left(\frac{y_0 r^{3/2}}{v_0\sqrt{D_R}}\right)
    \quad\text{where}~
    \mathcal{H}(w) =\frac{1-p(w)}{w^{2/3} p(w)}.
    \label{abp-scaling}
\end{align}

While the integral in \eref{p(w)-ABP-reset} is hard to evaluate exactly, the limiting behaviour of 
the function $p(w)$ can be found. For small $w$, the integral is dominated by large $z$ behaviours of the integrand. A naive large $z$ approximation of the integrand results in a divergence from around $z=0$. However, such divergence is absent (to the leading order) for the function $p'(w)$. Hence, we first evaluate $p'(w)$ with the large $z$ approximation of the integrand and then integrate with respect to $w$ to obtain
\begin{equation}
    p(w) = 1 -\frac{3^{5/6} w^{1/6}}{2^{1/6} \Gamma (1/3)}  -\frac{3^{1/3} w^{2/3} \Gamma (1/3)}{4 \Gamma (2/3)}+\dotsb\quad\text{as}~ w\to 0.
    \label{abp:smallw}
\end{equation}
where the integrand constant is fixed by the condition $p(0)=1$. For large $w$, the integral \eref{p(w)-ABP-reset} is dominated by the small $z$ behaviour of the integrated, i.e., $\text{Ai}(z^{-2/3})\left[1+ \frac 1{4 \sqrt{\pi}} \Gamma\left(-\frac 12, \frac 2{3z} \right)\right]\to e^{-\frac{2}{3 z}} z^{1/6}/(2 \sqrt{\pi })$ to the leading order, which results in,
\begin{equation}
    p(w) \to 
    \sqrt{\frac{3}{8}} e^{- \sqrt{8w/3}} \quad \text{as}~ w\to \infty.
    \label{abp:largew}
\end{equation}
Hence, ${\cal H}(w)$ diverges as $w^{-1/2}$ as $w \to 0$ and $e^{ \sqrt{8w/3}} $ as $w\to\infty$. Consequently, the mean first-passage time in the presence of resetting diverges as a power-law ($T_y\sim r^{-3/4}$) as $r\to 0$ and stretched exponentially ($\ln T_y \sim  r^{3/4}$) as $r\to \infty$. The function $p(w)$ is minimum at $w^*=0.358\dots$, and consequently, the mean-first passage time $T_y$ is minimum at the optimal resetting rate given by,
$\bigl[y_0/(v_0\sqrt{D_R})\bigr]^{2/3}\, r^* = 0.504\dots$. The scaling behavior in \eref{abp-scaling}, along with the asymptotic behaviors \eref{abp:smallw} and \eref{drabp:largew} is illustrated in Fig.~\ref{fig:drabp_reset_y}.

\begin{figure}
    \centering
    \includegraphics[width=7 cm]{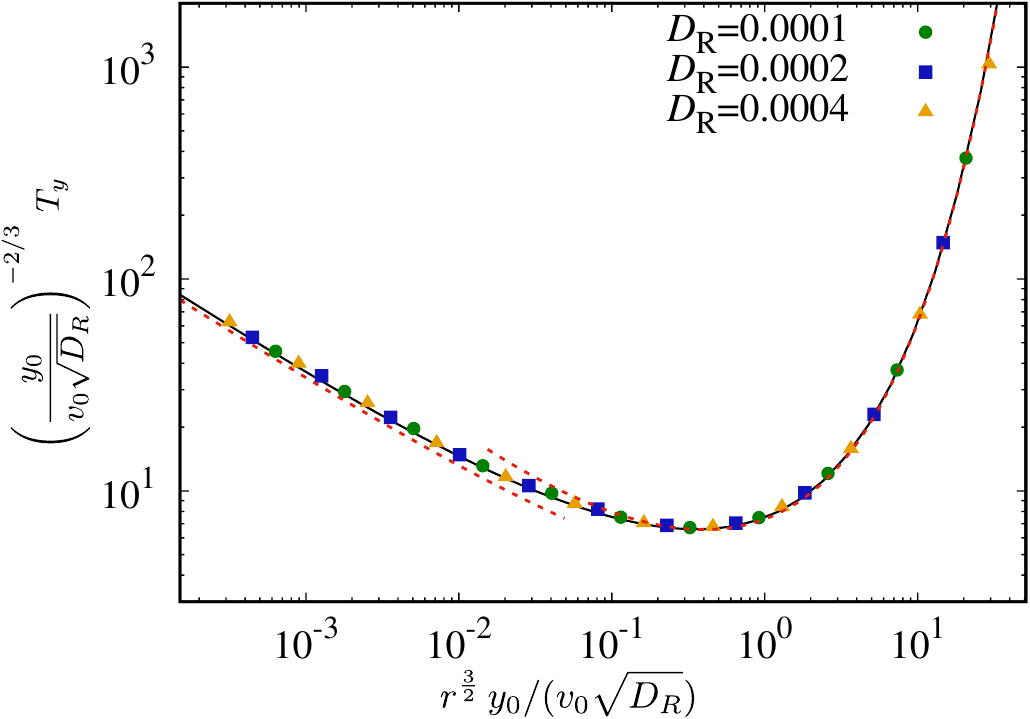}
    \caption{ABP under reset: Scaled MFPT corresponding to the $y$-component in the short-time regime $D_R \ll r $. The symbols indicate the data obtained from numerical simulations. The solid black line indicates the predicted scaling function \eref{abp-scaling}, while the dashed red lines correspond to the asymptotic expressions \eref{abp:largew} and \eref{abp:smallw}, of the scaling functions for small and large $w$.}
    \label{f:abp-scaled}
\end{figure}

Recently, the first-passage properties of an ABP confined within a circular absorbing boundary was investigated perturbatively around the passive limit~\cite{di2023active}, which was shown to depend more strongly on the magnitude of the propulsion velocity, compared to the rotational diffusion coefficient. First-passage properties of an ABP in the presence of Lorentz force and stochastic resetting was investigated in \cite{abdoli2021stochastic}, where it was shown that, in the presence of an inhomogeneous magnetic field, an active particle takes longer to reach a target compared to a passive particle.

\section{Direction reversing active Brownian particles}
\label{sec:drabp}

The two most common models of active motion discussed so far, namely, run-and-tumble and active Brownian particles, exhibit qualitatively different types of motion--- the RTP exhibits intermittent,  discrete changes of orientation, while the orientation for an ABP changes continuously. There exists a class of microorganisms, which shows an ABP-like motion with intermittent directional reversals. Examples include soil bacteria like Myxococcus xanthus~\cite{wu2009periodic,thutupalli2015directional,leonardy2008reversing,liu2019self}, Pseudomonas putida~\cite{harwood1989flagellation,theves2013bacterial}, Pseudomonas citronellolis~\cite{taylor1974reversal}, marine bacteria like Pseudoalteromonas haloplanktis, and other bacteria like Shewanella putrefaciens~\cite{johansen2002variability,barbara2003bacterial}. A minimal statistical model to describe such motion is the so-called the direction reversing active Brownian particle (DRABP) model~\cite{santra2021active}, which in two dimensions evolves by the Langevin equations,
\begin{align}
    \dot{x}(t) &=v_0 \sigma(t) \cos \theta(t), ~~ \dot{y}(t)= v_0 \sigma(t)\sin \theta(t), ~~\text{and}~~ \dot\theta(t) = \sqrt{2D_R}\, \zeta(t),
    \label{drabp:langevin}
\end{align}
where $\sigma(t)$ is a dichotomous noise which alternates between $\pm 1$ with a rate $\gamma$ and $\zeta(t)$ is a Gaussian white noise with zero mean and $\la\zeta(t)\zeta(t')\ra=\delta(t-t')$. The continuous evolution of  $\theta(t)$ gives rise to an ABP-like motion between two successive reversal events, indicated by $\sigma\to -\sigma$ transitions. 

The most interesting behavior is observed when the reversal time-scale $\gamma^{-1}$ is much smaller than the rotational time-scale $D_R^{-1}$. This leads to the emergence of three distinct dynamical regimes, namely, short-time regime $t \ll \gamma^{-1},D_R^{-1}$, intermediate-time regime $\gamma^{-1}\ll t \ll D_R^{-1}$ and long-time regime $t \gg \gamma^{-1},D_R^{-1}$. While in the long-time regime the DRABP behaves like a Brownian particle with an effective diffusion constant $D_{\text{eff}}=\frac{v_0^2}{2(D_R+2\gamma)}$, a strongly non-diffusive and anisotropic behavior is seen in the short and intermediate time regimes~\cite{santra2021active}.

The complex dynamical behavior of DRABP also leads to non-trivial first-passage properties with signatures of anisotropy. Similar to ABP, this anisotropic behavior is best characterized by looking at the survival probabilites along and orthogonal to the initial orientation. To this end, we look at the survival probabilities along $x$ and $y$ directions separately, by choosing $\theta(0)=0$ and $\sigma(0)=1$.

Let $Q_x(x_0,t)$ denote the probability that the DRABP, starting from the $x=x_0$ line, has not crossed the absorbing boundary at $x=0$. Similarly, $Q_y(y_0,t)$ denotes the survival probability for the $y$-component, starting from the $y=y_0$ line with an absorbing boundary at $y=0$. The long-time diffusive behavior of DRABP implies that the survival probabilities follow \eref{eq:Q_diff} with $D_{\text{eff}}=\frac{v_0^2}{2(D_R+2\gamma)}$.
On the other hand, in the short-time regime, the behavior of the DRABP is similar to that of an ABP, with $Q_x(x_0,t)\simeq 1$ and $Q_y(y_0,t)$ given by \eref{eq:Sy_RAP}.

 The interplay of the reversal and rotational diffusion leads to a novel first-passage behavior in the intermediate regime, which we discuss below. In this regime, the dichotomous noise $\sigma(t)$ emulates a Gaussian white noise $\xi(t)$ with zero mean and autocorrelation $\la\xi(t)\xi(t')\ra=\gamma^{-1}\delta(t-t')$. Thus the dynamics of DRABP in this regime is described effectively by the Langevin equations,
\begin{align}
    \dot{x}(t)=v_0\xi(t),\quad\dot{y}(t)=v_0\xi(t)\phi(t).
    \label{drabp:intermediate}
\end{align}
Clearly, the dynamics along the $x$-component is just a Brownian motion, leading to the survival probability given by \eref{eq:Q_diff} 
with an effective diffusion coefficient $v_0^2/(2\gamma)$.
The $y$-component, on the other hand, undergoes a diffusion with a stochastic diffusion coefficient, that itself undergoes a Brownian motion.

\begin{figure}
    \centering
    \includegraphics[width=\hsize]{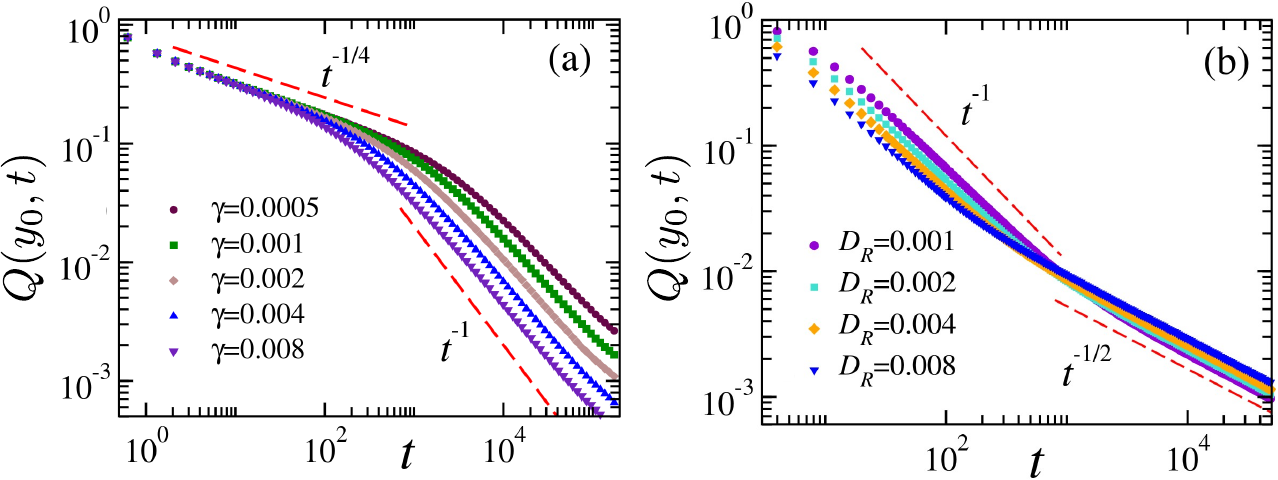}
    \caption{Survival probability of the $y$-component of the DRABP: (a) shows the crossover from the short-time $\sim t^{-1/4}$ behavior to the intermediate time $\sim t^{-1}$ behavior for $D_R=10^{-5}$ and $y_0=0.001$. (b) shows the crossover from the intermediate time $\sim t^{-1}$ to large-time $\sim t^{-1/2}$ decay for $\gamma=1$ and $y_0=0.1$.}
    \label{sy_drabp-free}
\end{figure}

The survival probability $Q_y(y_0,t)$ \ie , the probability that a particle starting with a $y(0)=0$, and  an initial orientation $\theta_0=0$ has not crossed the $y=0$ line till time $t$, is given by,
\bea
Q_y(y_0,t) = \int_0^\infty \id y ~ P(y,t;y_0), \label{eq:SP}
\eea
where $P(y,t;y_0)$ is the marginal probability distribution of the $y$-component in the presence of an absorbing wall at $y=0$, starting from the initial position $y_0.$ The corresponding forward Fokker-Planck equation for $P(y,\phi,t),$ \ie, the probability that  $y(t)=y$ and $\phi(t)=\phi,$ 
\bea
\frac{\partial }{\partial t} P(y,\phi,t) &=&\frac{v_0^2\phi^2}{2\gamma}\frac{\partial^2 }{\partial y^2} P(y,\phi,t) +D_R\frac{\partial^2 }{\partial \phi^2} P(y,\phi,t),
\label{fp_intermed}
\eea
with the initial condition $P(y,\phi,0)=\delta(y-y_0)\delta(\phi)$ and boundary conditions $P(y,\phi,t)\to 0$ as $\phi(t)\to\pm \infty$ and $P(0,\phi,t)=P(\infty,\phi,t)=0.$ Note that, we have suppressed the initial position dependence for notational convenience.
Making a change of variable $\frac {v_0^2t}{2\gamma} \to t$ and taking a sin-Laplace transform, defined by,
\bea
\tilde P(k,\phi,s) = \int_0^\infty dt~ e^{-st} \int_0^\infty dy~ \sin(k y) \, P(y,\phi,t),\quad\text{with  }k>0,
\eea
the Fokker-Planck equation \eref{fp_intermed} reduces to 
\bea
\Lambda^2\frac{\id^2 }{\id \phi^2} \tilde{P}(k,\phi,s)-(s+ \phi^2 k^2) \tilde{P}(k,\phi,s)= - \sin(ky_0)\delta(\phi),\label{eq:Ptdiff}
\eea
with $\Lambda^2=2\gamma D_R/v_0^2 $.
Using the boundary conditions $\tilde P(k,\phi,s) \to 0$ for $\phi \to \pm \infty,$ the continuity (and discontinuity) of $\tilde P(k,\phi,s)$ (and its derivative) across $\phi=0$, we get the exact solution,
\bea
\tilde{P}(k,\phi,s)&=& \frac{2^{\frac q 2} \sin(k y_0)}{\sqrt{8 \pi k \Lambda^3}} \Gamma\left(\frac q 2 \right) D_{-q}\left(|\phi| \sqrt{\frac{2k}\Lambda}\right).
\eea
where, $q= \frac 12(1+ \frac s{k \Lambda}).$ Since only $P(y,t)$ is needed to obtain the survival probability, we integrate over $\phi$ to get the sine-Laplace transform of $P(y,t;y_0)$,
\bea
\hat{P}(k,s)=\frac{2 \sin(k y_0)}{s + k\Lambda} ~  _2F_1\left(1,\frac{q+1}2,\frac {q+2}2,-1\right), 
\eea
where $_2F_1(a,b,c,z)$ denotes the Hypergeometric function \cite{NIST:DLMF}. This can be inverted exactly~\ref{sec:app}, to get a scaling form for the forward propagator,
\bea
P(y,t;y_0) = \frac{1}{4\Lambda t}\Bigg[ f\Big(\frac{y-y_0}{4\Lambda t} \Big) - f\Big(\frac{y+y_0}{4\Lambda t}\Big) \Bigg],
\eea
where, the scaling function $f(z)$ is given by,
\bea
f(z) &=& \frac 1\pi \int_0^\infty d\kappa~ \frac{\cos(\kappa z)}{\sqrt{\cosh (\kappa/2)}} 
= \frac 1{\sqrt {2 \pi^3}} \Gamma \Big(\frac 14 + iz \Big)\Gamma \Big(\frac 14 - iz \Big).
\label{scaling_f}
\eea
Thus, the survival probability, given by Eq.~\eqref{eq:SP}, also has a scaling form,
\bea
Q_y(y_0,t)=g\left(\frac {y_0}{v_0t}\sqrt{\frac \gamma{8 \dr}}\right),
\label{eq:S_scaling2}
\eea
where the scaling function $g(z_0)$ is given by,
\bea
g(z_0) = \int_0^\infty dz~ [f(z-z_0) -f(z+z_0)] = 2 \int_0^{z_0} dz~ f(z).
\label{g(z):def}
\eea
The large time behavior of the survival probability can be easily extracted by noting that $g(z_0)= 2 z_0 f(0) + O(z_0^2)$ for $z_0 \ll 1.$ Thus, we have,
\bea
Q_y(y_0,t)&\approx & \frac {\Gamma(1/4)^2}{2 \pi^{3/2}} \sqrt{\frac \gamma \dr} \, \frac {y_0}{v_0 t}\quad\text{for}\quad\frac{y_0}{v_0 t}\ll\sqrt{\frac{\dr}{\gamma}}\ll 1. 
\label{persis}
\eea
Thus the survival probability in the direction orthogonal to the initial orientation exhibits non-monotonic decay exponents, increasing from an initial $\alpha=1/4$ to $\alpha=1$ at around $t\sim O(\gamma^{-1})$, and then decreasing to $\alpha=1/2$ at $t\sim O(D_R^{-1})$ as showin in Fig.~\ref{sy_drabp-free}. \\

\noindent{\bf Survival Probability along the direction of the initial orientation}
The survival probability along the direction of the initial orientation, i.e., the probability that a particle starting from $x_0>0$ with an initial orientation $\theta_0=0$ has not crossed the line $x=0$ up to time $t$ is, however, less interesting.
 At very short times ($t\ll\gamma^{-1},\dr^{-1}$) the trajectories undergo none or very few reversals, as a result the particle starting with the initial orientation $\theta_0=0$ always moves away from the line $x=0$. Due to this the particle almost always survives and the corresponding survival probability remains unity. In the intermediate and large-time regime the particle exhibits diffusive motion as a result of which survival probability falls off as $t^{-1/2}$.

\subsection{DRABP with resetting}
Similar to RTP and ABP, the introduction of stochastic resetting helps a DRABP in finding a target in a finite amount of time. In the following, we discuss the first-passage properties of a DRABP along the direction orthogonal to the initial orientation, which shows a very rich behavior even in the absence of resetting.
We choose initial conditions $\theta(0)=0$ and $\sigma(0)=1$, and investigate the survival probability of the $y$-component, starting from $y=y_0$, in the presence of an absorbing wall at $y=0$. At each resetting event, the position and orientation are both reset to their initial values at a constant rate $r$. In this case, the renewal equation for the survival probability $Q^{(r)}(y_0,t)$ is the same as \eref{abp:renewal}. Consequently the mean-first passage time can also be determined from the Laplace transform of the survival
probability in the absence of resetting using \eref{abp:mfpt}.

\begin{figure}
    \centering    \includegraphics[width=7 cm]{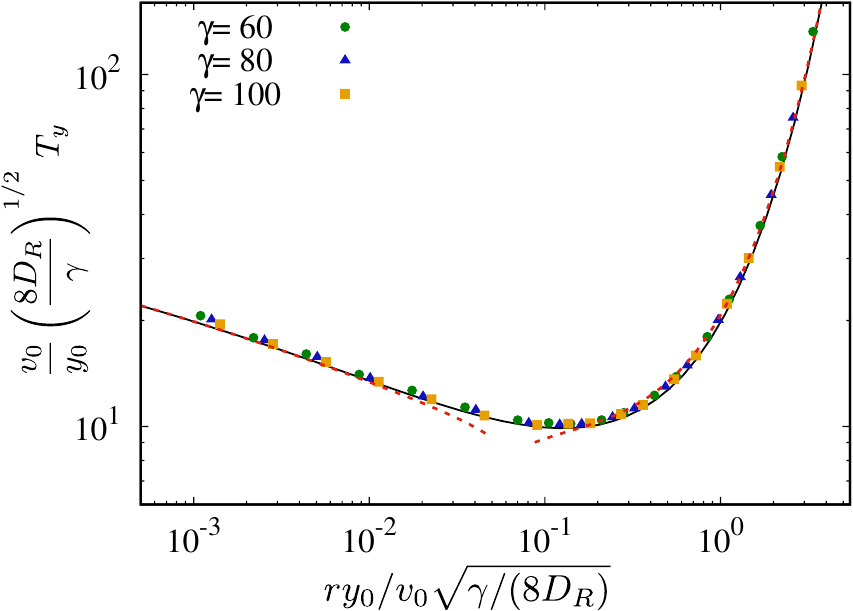}
    \caption{DRABP under reset: Scaled MFPT corresponding to the $y$-component in the intermediate regime $D_R \ll r \ll \gamma$. The symbols indicate the data obtained from numerical simulations. The solid black line indicates the predicted scaling function \eref{drabp:scaling}, while the dashed red lines correspond to the asymptotic expressions \eref{drabp:largew} and \eref{drabp:smallw}, of the scaling functions for small and large $w$.}
\label{fig:drabp_reset_y}
\end{figure}

If the resetting rate $r$ is much larger compared to $D_R$ and $\gamma$, the dynamics of the $y$-component between successive resetting events are well approximated by an RAP. Consequently, the mean first-passage time shows the same scaling behavior as \eref{abp-scaling}. On the other hand, a different scenario emerges when $\gamma\gg r \gg D_R$, where the $y$-process, between two resetting events, is effectively described by \eref{drabp:intermediate}. Hence, to determine the mean first-passage time, using \eref{abp:mfpt}, we require the Laplace transform of the survival probability \eref{eq:S_scaling2}, which is given by,
\begin{align}
    \tilde Q(y_0,r)
   \equiv\int_0^\infty e^{-rt} Q(y_0,t)\, dt
    =\frac{1}{r}\left[ 1-p\left(\frac{r y_0}{v_0}\sqrt{\frac{\gamma}{8D_R}}\right)\right], 
\end{align}
where
\begin{equation}
p(w)=2\int_0^{\infty}  e^{-w /z}f(z)\, dz.
\label{eq:p(w)drabp}
\end{equation}
The MFPT, in turn, has a scaling form,
\bea 
T_y = \frac{y_0}{v_0} \sqrt{\frac{\gamma}{8 D_R}}\, {\cal G}\left( \frac{r y_0}{v_0} \sqrt{\frac{\gamma}{8 D_R}} \right),
\eea 
with the scaling function,
\begin{align}
    {\cal G}(w) = \frac{1- p(w)}{\, w \, p(w)}.
    \label{drabp:scaling}
\end{align}
 This analytical scaling form is compared with numerical simulations in Fig.~\ref{fig:drabp_reset_y}, where a perfect scaling collapse of the MFPT for different values of $\gamma$ is observed.

Although it is hard to find closed form expressions for $p(w)$, its behavior for small and large $w$ can be extracted. To find the leading small-$w$ behavior, it is convenient first to consider the second derivative $p''(w)=2 w^{-1}\int_0^\infty F_1(z,w)\, f(z)\, dz$, where $F_1(z,w)=w e^{-w/z}/z^2$ is a Fr\'echet distribution. In the limit $w\to 0$, most of the weight of $F_1(z,w)$ is concentrated around $z=w$, where $f(w)$ to the leading order can be replaced by $f(0)$.  Hence, $p''(w) \approx 2 f(0)/w$ to the leading order, where we have used the normalization $\int_0^\infty F_1(z,w)\, dz =1$. This becomes more apparent when one enlarges the region of integration near $z=0$ by a change of variable $z=e^u$, which 
 yields $p''(w) = 2 w^{-1} \int_{-\infty}^\infty G(u - \ln w)\, f(e^u)\, du$, where $G(x) = e^{-x - e^{-x}}$ is the well-known Gumbel distribution.  In the limit $w\to 0$, the distribution $G(u-\ln w)$ is peaked around $u=\ln w \ll 0$, where $f(e^u)$ is almost flat and to the leading order, can be approximated by its limiting value $f(0)$ for $u\to -\infty$. Therefore, using $\int_{-\infty}^\infty G(x)\, dx =1$,  for small $w$, to the leading order, we have $p''(w) \approx 2 f(0)/w$. Integrating twice, to the leading order, we get $p(w) \approx 2 f(0) w
\ln w + 1$, where the integration constant $1$ is fixed by the exact result $p(0)=1$. In fact, one can go beyond this heuristic argument and find $p(w)$ for small $w$, to any order, systematically. We refer to \cite{santra2022effect}, where, from exact calculation, we found the integral $-p'(w)=-a_1\ln w+b_1+c_1w+d_1w^2+e_1w^2\ln w+h_1w^3+O(w^4)$, where $a_1=2f(0)\approx3.34\dots$, $b_1\approx-6.93\dots$, $c_1\approx22.25\dots$, $d_1\approx13.47\dots$, $e_1\approx28.71\dots$, and $h_1= -60.29\dots$. Integrating once, we get,
\begin{align}
p(w) = 1 &+ \left(a_1 -\frac{e_1 w^2}{3}\right) w \ln w  -(a_1+b_1) w-\frac{c_1 w^2}{2} \cr
+& {\Big(\frac{e_1}{9}-\frac{d_1}{3}\Big)} w^3-\frac{h_1 w^4}{4} + O(w^5),
\label{drabp:smallw}
\end{align}
 where, once again, we fixed the integration constant using $p(0)=1$.

Due to the presence of the essential singularity $e^{-w/z}$ in \eref{eq:p(w)drabp}, the  behaviour of the integral for large $w$ is dominated by large-$z$ behaviour of $f(z)$. Hence, using the asymptotic behavior  $f(z)= e^{-\pi z}\sqrt{\frac{2}{\pi z}}\Bigl[1+(8z)^{-2} + \dotsb \Bigr]$ for $z \gg 1$, in \eref{eq:p(w)drabp} and performing the integral, 
we get,
\begin{align}
   p(w) = \frac{e^{-2 \sqrt{\pi w}}}{\sqrt{2\pi}} \left[4+ \frac{\pi}{16 w}+ \frac{\sqrt{\pi}}{32 w^{3/2}} + O\bigl(w^{-2}\bigr) \right].
   \label{drabp:largew}
\end{align}

Therefore, the mean first-passage time in the presence of resetting diverges logarithmically  ($T_y\sim -\ln r$) as $r
\to 0$ and stretched exponentially ($\ln T_y \sim \sqrt{r}$) as $r\to \infty$. At an optimal resetting rate given by 
$\sqrt{\gamma/(8 D_R)}\, (y_0/v_0)\,  r^*= 0.2656\dots$, $T_y$ is minimum. The predicted asymptotic behaviors \eref{drabp:smallw} and \eref{drabp:largew} of the scaling function is also compared with the scaled MFPT, obtained from numerical simulations, in Fig.~\ref{fig:drabp_reset_y}, which show excellent match.

\section{Conclusion and outlook}

In this chapter, we have reviewed recent theoretical results, and reported some new findings about the first-passage properties of self-propelled/active particles. The dynamics of these particles are modeled by overdamped Langevin equations with colored noises, the correlation-times of the noise being a measure of the activity. We discuss the first-passage and survival probabilities of the three most well-known active particle models, namely RTP, ABP and DRABP, in the presence of an absorbing target. While all these processes eventually become diffusive at times much larger than the active time-scale, leading to the familiar $\sim t^{-3/2}$ tail of the first-passage time distribution, the presence of the activity gives rise to a set of interesting scaling behavior and non-trivial persistence exponents. 
We also review how stochastic resetting expedites the search process by an active particle. Because of the presence of an additional internal orientation, these particles exhibit a rich first-passage behaviour in the presence of resetting. In particular, we obtain interesting non-trivial functional dependence of the mean first-passage time on the resetting rate $r$ for the different models.

The advancement in single-particle tracking experiments provides an excellent opportunity to extract model parameters for real systems. The theoretical results, reviewed here, would then aid in the design of artificial micro/nanobots, that are increasingly employed in targeted drug delivery to remote intracellular sites~\cite{xin2021environmentally}.

Apart from the potential applications, there are several open questions which are of interest from the theoretical point of view. The exponential waiting time between two successive tumblings or directional reversals, considered here, is an idealization of the real phenomena~\cite{theves2013bacterial}, and it would be interesting to study how the first-passage properties are altered when the waiting times are selected from distributions that are non-Poissonian~\cite{grossmann2016diffusion,detcheverry2017,zaburdaev2015levy}.
 Active particles quite often have a `chirality', which refers to a constant drift in the evolution of the orientation~\cite{liebchen2022chiral,das2023chirality}. The first-passage properties of chiral active particles are virtually unexplored. 
For ABP and DRABP, we considered a complete reset, i.e., both position and orientation being reset to the initial condition. It would be interesting to consider more general resetting protocols, such as, partial resetting~\cite{santra2020run,kumar2020active} and non-instantaneous resetting~\cite{gupta2020stochastic,santra2021brownian,radice2021one}.

A particularly challenging question is how the first-passage properties change in higher dimensions. For example, considering specific target zones (say, a circular hole at the origin) in two-dimensions, instead of a line target, which has been discussed in this review. This is particularly hard for active particles as the absorbing boundary conditions typically involve the orientation of the particle as well, unlike their passive Brownian counterpart~\cite{spitzer1991some}. 
Last but not the least, considering that the active particles, modeling living organisms, are typically found in groups or colonies with competition for resources, exploring the first-passage properties of interacting active particles, especially in the presence of passive crowders~\cite{biswas2020first, khatami2016active}, could be a promising avenue for future research.

\section{Appendix}
\label{sec:app}
In this section, we provide the detailed calculation of the survival probability of the DRABP in the intermediate regime starting from \eref{eq:Ptdiff}. For $\phi\ne 0,$ the general solution of Eq.~\eqref{eq:Ptdiff} is given by
\bea
\tilde{P}(k,\phi,s)=a~ D_{-q}\left(\phi \sqrt{\frac{2k}\Lambda}\right)+b~ D_{-q}\left(-\phi \sqrt{\frac{2k}\Lambda}\right)
\eea
where $q=\frac 12 (1 +\frac{s}{k\Lambda})$, $D_\nu(z)$ denotes the parabolic cylinder function \cite{NIST:DLMF} and $a,b$ are two arbitrary constants independent of $ \phi$. Using the boundary conditions for $\phi \to\pm\infty$, and the fact that $\tilde{P}(k,\phi,s)$ is continuous at $\phi=0$  we have,
\bea
\tilde{P}(k,\phi,s)= \left \{
\begin{array}{ll}
     a~ D_{-q}\left(\phi \sqrt{\frac{2k}\Lambda}\right) , & \text{for  } \phi >0 \\
      a~ D_{-q}\left(-\phi \sqrt{\frac{2k}\Lambda}\right), & \text{for } \phi <0.
    \end{array} \right.
    \label{homosol}
\eea
Integrating Eq.~\eqref{eq:Ptdiff} across $\phi=0,$ we get, 
\bea
\frac{\id \tilde{P}}{\id \phi}\bigg|_{\phi=0^+}-\frac{\id \tilde{P}}{\id \phi}\bigg|_{\phi=0^-}=-\frac{\sin(k y_0)}{\Lambda^2}.\n
\eea
Using this equation with Eq.~\eqref{homosol} we get, 
\bea
a= \frac{2^{\frac q 2} \sin (k y_0)}{\sqrt{8 \pi k \Lambda^3}} \Gamma\left(\frac q 2 \right).
\label{eq:a}
\eea
Finally, combining Eq.~\eqref{eq:a} with Eq.~\eqref{homosol} we get,
\bea
\tilde{P}(k,\phi,s)&=& \frac{2^{\frac q 2} \sin(k y_0)}{\sqrt{8 \pi k \Lambda^3}} \Gamma\left(\frac q 2 \right) D_{-q}\left(|\phi| \sqrt{\frac{2k}\Lambda}\right),
\eea
where, as before, we have denoted $q= \frac 12(1+ \frac s{k \Lambda}).$ 
Since we are interested in the $y$-marginal distribution, we integrate over $\phi$ to get,
\bea
\hat{P}(k,s)=\frac{2 \sin(k y_0)}{s + k\Lambda} ~  _2F_1\left(1,\frac{q+1}2,\frac {q+2}2,-1\right) \n
\eea
which is the  $\sin$--Laplace transform of $P(y,t).$
Here $_2F_1(a,b,c,z)$ denotes the Hypergeometric function \cite{NIST:DLMF}. 

To find the position distribution we need to invert the  Laplace and $\sin$ transformations. The inverse Laplace transform is defined by the integral,
\bea
\hat P(k,t) = \int_{c_0-i \infty}^{c_0+ i \infty} \id s~ e^{st} \tilde{P}(k,s)
\eea
where $c_0$ is chosen such that all the singularities of the integrand lie to the left of the $Re(s)=c_0$ line. To compute the above integral let us first recast $\tilde P(k,s)$ as,
\bea
\tilde{P}(k,s) &=& \frac{2  \sin(k y_0)}{s + k\Lambda} ~  _2\tilde F_1\left(1,\frac{q+1}2,\frac {q+2}2,-1\right) \Gamma\left(\frac {q+2}2 \right), \label{eq:Pks}
\eea
where $_2\tilde F_1(a,b,c,z) = _2F_1(a,b,c,z)/\Gamma(c)$ denotes the regularized Hypergeometric function which is analytic for all values of $a,b,c$ and $z.$ From Eq.~\eqref{eq:Pks}, it is straightforward to identify the singularities of $\tilde{P}(k,s),$ on the complex $s$-plane all of which lie on the negative real $s$-axis: $s_n=-k \Lambda(4n+5)$ with $n=-1,0,1,2,\cdots$  where  $s_{-1}$ comes from the prefactor $(s+\Lambda k)^{-1}$ while $s_{n \ge 0}$ are obtained from the singularities $q_n=-2(n+1)$  of  $\Gamma\left(\frac{q+2}2 \right).$

The inverse Laplace transform of Eq.~\eqref{eq:Pks} can then be expressed as
\bea
\hat P(k,t) &=& \sum_{n=-1}^{\infty} e^{s_n  t} R_n, \label{eq:res_sum}
\eea
where $R_n$ denotes the  residue of $\tilde{P}(k,s)$ at $s=s_n.$ These residues can be computed exactly and turn out to be 
\bea
R_n =  2 \sin(k y_0) \frac{(-1)^{n+1}}{(n+1)!} ~_2\tilde F_1 \left(1,-n - \frac 12,-n,-1 \right). \n
\eea
Using the above expression in Eq.~\eqref{eq:res_sum} and shifting $n \to n-1$, we get,  
\begin{equation}
\hat P(k,t) = 2 \sin(k y_0) \sum_{n=0}^{\infty} \frac{(-1)^n}{n!}  e^{- (1+4 n) k \Lambda t}  2\tilde F_1 \left(1, -n+ \frac 12, -n+1,-1 \right).\label{eq:Pkt1}
\end{equation}
Using properties of Hypergeometric functions, it can be shown that 
\bea
_2\tilde F_1\left(1, -n+ \frac 12, -n+1,-1 \right) = \frac {(-1)^n}{\sqrt 2} \left({-1/2 \atop n} \right) n! ~~\n
\eea
Substituting the above identity in Eq.~\eqref{eq:Pkt1} we finally get,
\bea
\hat{P}(k,t)&=& \sqrt{2} \sin(k y_0) e^{-k \Lambda  t}\sum_{n=0}^{\infty} \left({- \frac 12 \atop n} \right) e^{- 4 n k \Lambda  t } 
= \frac{\sin(k y_0)}{\sqrt{\cosh \left(2 k  \Lambda t \right)}} \label{eq:Linv}
\eea
The position distribution is given by the inverse $\sin$-transform,
\bea
P(y,t;y_0) =   \frac 2 \pi \int_{0}^\infty dk ~\sin (k y) \hat P(k,t) &=& \frac 1 \pi \int_{0}^\infty dk ~ \frac {2 \sin (k y) \sin (k y_0)}{\sqrt{\cosh \left(2 k  \Lambda t \right)}}. ~~
\eea
Using the trigonometric identity $2 \sin (k y) \sin (k y_0)=[\cos (k(y-y_0)) - \cos (k(y+y_0)) ]$, it is straightforward to see that $P(y,t)$ has a scaling form,
\bea
P(y,t;y_0) = \frac{1}{4\Lambda t}\Bigg[ f\Big(\frac{y-y_0}{4\Lambda t} \Big) - f\Big(\frac{y+y_0}{4\Lambda t}\Big) \Bigg],\n
\eea
where, the scaling function $f(z)$ can be evaluated exactly,
\bea
f(z) &=& \frac 1\pi \int_0^\infty d\kappa~ \frac{\cos(\kappa z)}{\sqrt{\cosh (\kappa/2)}} 
= \frac 1{\sqrt {2 \pi^3}} \Gamma \Big(\frac 14 + iz \Big)\Gamma \Big(\frac 14 - iz \Big).
\label{scaling_f}
\eea
The survival probability, given by Eq.~\eqref{eq:SP}, also has a scaling form,
\bea
S_y(t;y_0) = g\left(\frac {y_0}{4 \Lambda t}\right), \label{eq:S_scaling}
\eea
where $g(z_0)$ is given by,
\bea
g(z_0) = \int_0^\infty dz~ [f(z-z_0) -f(z+z_0)] = 2 \int_0^{z_0} dz~ f(z).
\label{g(z):def}
\eea
In terms of the original notation $t \to v_0^2 t /(2 \gamma)$, and $\Lambda^2=2\gamma D_R/v_0^2,$
\bea
S_y(t;y_0)=g\left(\frac {y_0}{v_0t}\sqrt{\frac \gamma{8 \dr}}\right).
\label{eq:S_scaling2}
\eea
The large time behavior can be extracted easily by taking $z_0 \ll 1,$
\bea
g(z_0)= 2 z_0 f(0) + O(z_0^2).
\eea
Thus, we have,
\bea
S(y_0,t)&\approx & \frac {\Gamma(1/4)^2}{2 \pi^{3/2}} \sqrt{\frac \gamma \dr} \, \frac {y_0}{v_0 t}\quad\text{for}\quad\frac{y_0}{v_0 t}\ll\sqrt{\frac{\dr}{\gamma}}\ll 1. 
\label{persis}
\eea
Using this result we conclude in the main text that the survival probability of a DRABP in the time regime (II) $\gamma^{-1}\ll t\ll D_R^{-1}$ has a power-law decay with persistence exponent $\alpha_y=1.$

Note that, the exact first-passage distribution $F_y(t)= -\partial_t S_y(t;y_0)$ can be easily  computed from Eq.~\eqref{eq:S_scaling2},
\bea
F_y(t;y_0)=\frac{y_0\, \sqrt{2} \gamma^{3/2}}{v_0^3\, t^2\sqrt {\dr}} f\left(\frac {y_0}{v_0t}\sqrt{\frac \gamma{8 \dr}}\right).
\label{fpt_exact}
\eea 
In terms of the scaled time $\tau=v_0 t\sqrt{\frac{8\dr}{\gamma}}$, 
\bea
S_y(\tau)=g(1/\tau)=2\int_{0}^{1/\tau}dz f(z)=2\int_{\tau}^{\infty}\frac{dz}{z^2}\, f(1/z)
\label{surv_scaled}
\eea
where $f(z)$ is defined in Eq.~\eqref{scaling_f}.

\bibliographystyle{spmpsci}  
\bibliography{ref}

\end{document}